            \documentclass[12pt]{article}
\usepackage{amsfonts}
\usepackage{amssymb}

\makeatletter
\renewcommand \thesection {\@arabic\c@section}
\renewcommand\thesubsection   {\thesection.\@arabic\c@subsection}
\renewcommand\thesubsubsection{\thesubsection .\@arabic\c@subsubsection}
\renewcommand\theparagraph    {\thesubsubsection.\@arabic\c@paragraph}
\renewcommand\section{\@startsection {section}{1}{\z@}%
                                   {-3.5ex \@plus -1ex \@minus -.2ex}%
                                   {1.9ex \@plus.2ex}%
                                   {\normalfont\large\bfseries\centering}}
\renewcommand\subsection{\@startsection{subsection}{2}{\z@}%
                                     {-2ex\@plus -1ex \@minus -.2ex}%
                                     {1.2ex \@plus .2ex}%
                                    {\normalfont\normalsize\bfseries\centering}}
\renewcommand\subsubsection{\@startsection{subsubsection}{3}{\z@}%
                                     {-2ex\@plus -1ex \@minus -.2ex}%
                                     {.5ex \@plus .2ex}%
                                     {\normalfont\normalsize\em}}
\renewcommand\paragraph{\@startsection{paragraph}{4}{\z@}%
                                    {3.25ex \@plus1ex \@minus.2ex}%
                                    {-1em}%
                                    {\normalfont\normalsize\em}}
\renewcommand\subparagraph{\@startsection{subparagraph}{5}{\parindent}%
                                       {3.25ex \@plus1ex \@minus .2ex}%
                                       {-1em}%
                                      {\normalfont\normalsize\em}}
\makeatother

\newcommand{\Rset}{\mathbb{R}}

\begin{document}

\title{\textbf{Book Review of Jean Bricmont's \\ ``Making Sense of Quantum Mechanics''\\
 {\small{\textbf{Springer, 2016}}}}\vspace{-10pt}}
\author{\sc{Michael K.-H. Kiessling}\\ \vspace{-5pt}
	{\small{Department of Mathematics}}\\ \vspace{-5pt}
	{\small{Rutgers, The State University of New Jersey}}\\ \vspace{-5pt}
	{\small{110 Frelinghuysen Rd., Piscataway, NJ 08854}}}
\date{\hrule
\smallskip
\noindent
\footnotesize{
\copyright{2017} The author.  Reproduction of this review,
for noncommercial purposes only, is permitted.}\smallskip
\hrule}
\maketitle\vspace{-15pt}

\centerline{\hfill\small ``Given that quantum mechanics was discovered ninety years ago, the}

\centerline{\hfill \small present rather low level of understanding of its deeper meaning may}

\centerline{\hfill \small be seen to represent some kind of intellectual scandal.'' \cite{ETH} $\;\qquad$ }
\smallskip

\noindent
 Folklore has it that physicists, when they get ``too old'' for doing ``real physics,''
turn into ``hobby philosophers'' pondering the elusive ``deeper meaning''
of quantum mechanics, very much to the bemusement of their more practically minded younger peers.
 Closer to the truth than such feel-good folklore is the realization that physicists who try to
understand what our practically successful most fundamental theories\footnote{Beside quantum 
  mechanics, I put general relativity theory in this category.}
say about our world are also well aware of the professional mine field on which they 
are traipsing.
 Thus, with some notable exceptions, they simply don't dare to risk ruining their scientific
reputation with an offering of their secret thoughts on the subject before reaching retirement age.
 Even then, with ones career no longer in danger, such a step requires courage --- after all,
who enjoys the thought of ``gossip among physicists brand[ing one] {`senile'}.''\footnote{I have
adapted this quote from p.281 of the book, where Bricmont quotes John Clauser (his ref.[97]) saying
furthermore: 
``I was personally told as student that these men [Einstein, Schr\"odinger, de Broglie] had become
senile, and that clearly their opinions could no longer be trusted in this regard. This gossip was
repeated to me by a large number of well-known physicists from many different prestigious institutions.
... Under the stigma's unspoken `rules,' the worst sin that one might commit was to follow Einstein's
teaching and to search for an explanation of quantum mechanics in terms of hidden variables, as Bohm
and de Broglie did.''}
 But for Jean Bricmont, well-known not only to mathematical physicists for his contributions, many jointly with Antti Kupiainen, 
to the renormalization group,
but also to a wider audience for his sharp-tongued and witty criticism,
jointly with Alan Sokal, of postmodern ``non-science'' \cite{SokalBricmont},
the quest for understanding was too important to wait that long. 
 He took the unusual step and opted for early retirement while still in his prime, so that
he could focus his attention on writing this book about how to make sense of quantum mechanics,
undistracted and uninterrupted by academic duties.
 The outcome is a superbly written monograph which in terms of logical clarity, intellectual depth, 
and scholarship is without equal in its field.

 The first half of Bricmont's book is devoted to explaining the \emph{central conceptual problem} of 
orthodox textbook quantum mechanics (QM): it ``suffers from'' what the great mathematical physicist Arthur Wightman called 
``the malaise of the measurement problem'' \cite{wightmanPUB}.
 The name has historical reasons and does not in itself reveal what the problem is.
 Picking up on Wightman's quasi-medical lingo, a more appropriate diagnosis of QM's ailment would be 
``reality deficiency disorder.''

 For instance, (quoting from chapter 1)
Werner Heisenberg (p.5)\footnote{All references to page numbers refer to the book under review.}
 concluded that in QM 
``... the idea of an objective real world whose smallest parts exist objectively in the same sense in which stones 
and trees exist ... is impossible.''  
 John Wheeler in his theory of the ``participatory universe'' (mentioned on p.8) even suggested
that the present observation of the universe through our telescopes creates the fact that the whole universe had a real past.
 More recently, Anton Zeilinger (quoted on p.10) ``suggest[ed] that ... the distinction between 
reality and our knowledge of reality, between reality and information, cannot be made.''

 Hm.\footnote{This comment is inspired by --- in fact: copied from --- a recent New York Times op-ed piece by the great economist 
   Paul Krugman, where it was made in response to a different reality dismissal.\vspace{-10pt}}

 Denials of an objective reality are, in the opinion of the author, and also of the reviewer,
aberrations of human thought, plain and simple.
 In the case at hand they represent an outgrowth of several of QM's founding fathers', or defenders', denial of their own failure 
to understand what is `really going on' in the experiments, according to their theory: If `nothing real is going on,' then 
nothing needs to be understood --- problem `solved' --- but at what price?
 ``They behave[d] like people celebrating defeats as if they were victories'' (p.287).
 The claims by Bohr, Heisenberg, and others that \emph{any deeper explanation is impossible} only shows 
``the hubris of the scientists, claiming that their understanding of quantum mechanics was somehow an 
ultimate understanding'' (p.287).
 (Both quotes are from the last section of the book.)

 Be that as it may, what exactly did perplex these brilliant scientists so much that it enticed
them to make such radical proposals? 
  Bricmont, with his talent for explaining complicated things in a logically clear and compelling manner
by stripping them down to their essentials without oversimplifying, identifies two key ingredients of QM 
(Bricmont calls them ``mysteries'')
which have no counterpart in classical\footnote{More accurately, non-quantum physics.}  physics:
``interference and superposition (of the fundamental state)'' (chapter 2), 
and ``(Einstein) nonlocality''(chapter 4). 
 Inserted in between (and appropriately so!) is a chapter, called ``Philosophical Intermezzo,'' 
on what it means to do science. 

 As to the first ``mystery,''
the fundamental state of a material $N$-body system in, say, Newtonian physics is given by the positions and momenta of all the
$N$ (point) particles (relative to an inertial reference frame).
 A linear superposition of two such states is not a valid fundamental state in Newtonian physics.
 In colloquial language, you are either here or over there.
 It makes no sense to claim you were 32 $\%$ here and 68 $\%$ over there.
 But one could claim, for instance, that on 100 different occasions,
on 32 $\%$ of them you were here, and on 68 $\%$ of them over there.
 In other words, such a superposition represents an \emph{ensemble of fundamental states} in Newtonian physics.
 By contrast, in (non-relativistic) QM, the fundamental quantum state of an $N$-body system is its
Schr\"odinger wave function, or its Pauli spinor wave function if the spin of electrons or other fermions is taken
into account.
 Its time evolution, between ``measurements,'' is unitary and given by Schr\"odinger's or Pauli's linear wave equation, respectively.
 What's more, a linear superposition of two fundamental quantum states (wave functions) is again a fundamental
quantum state evolving according to the same unitary dynamics\footnote{I have on occasion heard the criticism that
 the linearity of the unitary quantum evolution is not a distinctive feature of QM because
 Newtonian $N$-point mechanics is equivalent to the evolution by the linear Liouville equation, and therefore, like 
  QM, ``a linear theory.'' 
  However, such claims 
overlook that the equivalence only holds for the fundamental state, which
 is represented by a Dirac delta function concentrated on a single point in $N$-particle phase space --- the linear
 superposition of two such Dirac delta functions is a state which is \emph{not} supported at a single point and therefore
 not a legitimate fundamental state of Newtonian $N$-point mechanics.}
between ``measurements.''
 Bricmont uses the example of a Mach--Zehnder interferometer to explain in simple terms how this can lead to 
paradoxical phenomena which have no counterpart in classical physics.

 In such situations the usual ``quantum-speak'' is to say things like ``a quantum particle'' (recall that by
\emph{particle} one conventionally means a localized microscopic object)
``travels two different paths at the same time,'' or such. 
 But instead of admitting that such absurd conclusions suggest that 
\emph{maybe the quantum state} (the wave function, or the density matrix if one prefers) \emph{is not all there is} 
for a ``quantum system,'' 
to avoid such absurd realities Bohr, Heisenberg, and other founding members of the so-called 
``Copenhagen interpretation of QM'' rather prefered to deny the existence of an objective reality at the microscopic
level unless the act of ``measurement'' by a real ``observer'' with a real ``classical device'' creates its reality. 

 To illustrate how absurd such claims are Schr\"odinger pointed out that the supposedly not directly experiencable weirdness
of the microscopic quantum level can easily be
magnified to the macroscopic everyday level by putting a cat into the story, which features also in Bricmont's chapter 2.
 The lamentable cat being both dead and alive at the same time is of course never observed in the real world,
but the conclusion that the cat was never in such a limbo to begin with is not allowed if one \emph{insists}, as done 
in orthodox QM, that the fundamental quantum state ``is all there is'' for ``a quantum system.'' 
 But then one is forced to conclude that the originally very real (and alive) cat, which was put into the box together 
with a radioactive nucleus and the detector coupled to (in Schr\"odinger's words) ``a hell machine'' which will kill the
cat when detecting the radioactive decay, this cat will ``cease to be real'' upon sealing the box, only to ``become
real'' again upon reopening the box to ``measure'' whether the radioactive nucleus has decayed and the cat is dead, or
has not decayed and the cat is still alive.
 Schr\"odinger emphatically insisted that \emph{this is absurd}, preposterous. 
 (Incidentally, many physicists who reject the denial of an objective microscopic reality by the adherents of the
``Copenhagen interpretation'' but hold on to the idea that the quantum state ``yields the complete description'' 
try to extricate themselves
from the dilemma by insisting that ``all conflicting scenarios are simultaneously real,'' but each one in a ``different world,''
without mutual interactions between the worlds once ``they are created'' in ``measurement-like situations.''
  Bricmont addresses various ``Many-Worlds Interpretations,'' a notion prominently associated with Hugh Everett III,
in chapter~6.)
 
 The classical enlightenment period of physics had assigned human beings roles of essentially ``passive observers of an 
objectively real universe unfolding in front of their eyes.'' 
 Of course, a perfectly passive observer (in classical physics, poetical licence for a technical device which quantifies
a quality of an object without disturbing it, registers the result, which is then read by a human without changing the result) 
is a fiction comparable to the notion of a ``test particle,'' yet it 
is an idealization which in the framework of classical physics is in principle analyzable 
as an interaction between an experimental setup and a physical object whose qualities 
are being quantified by the measurement --- a good approximation, with controllable error terms.
 By contrast, orthodox textbook QM (and quantum theory in general) invokes the notion of ``measurement'' at a fundamental level,
formalized mathematically by Johann von Neumann in his ``measurement axioms'' \cite{JvN}, 
though without stipulating what precisely constitutes ``a measurement'' --- very much in line with 
Niels Bohr's insistance that in quantum physics the act of observation / measurement is fundamental 
and in principle not further analyzable in terms of more fundamental processes.  

 But how could a ``measurement,'' i.e.
the quantification of a quality of an object (if the word ``measurement'' really means what it stands for, linguistically)  
not be analyzable in terms of events and processes involving these objects, and the objects which constitute the 
measuring device ---  at all? 
 As Bricmont explains, typically ``a measurement in QM does not \emph{measure} anything'' 
in the proper sense of the word, and similarly ``making an observation in QM usually
does not mean we are \emph{observing} anything.'' 
 And then, if physicists when they say ``measurement'' do not actually mean ``measurement in the sense of quantification
of a quality of an object,'' the objection to Bohr's pronouncement does not apply.
 ``Measurement'' and ``observation'' are unfortunate choices of terminology in QM which have contributed their ample
share to the confusion which befuddles the theory. 
 Yet if we replace them, as John Bell has proposed, 
with a more neutral and  more appropriate terminology such as ``experiment'' and ``registration of a result,'' 
a crucial deficiency remains: 
QM fails to offer an objective explanation of \emph{how} experiments could have any results at all.\footnote{Wightman put it thus: 
``Where do the facts come from?''\vspace{-10pt}} 
  
 Some, and possibly even many physicists would strongly disagree with the claim that this is a deficiency of QM. 
 If these physicists would basically agree on why the lack of an objective explanation is not a deficiency of QM,
there possibly could be a serious issue to contemplate.
 However, as Steven Weinberg wrote recently \cite{Weinberg}: ``It is a bad sign that those physicists today who 
are most comfortable with quantum mechanics do not agree with one another about what it all means.''
 The reason, it seems, is that quantum physicists disagree to an astonishing degree on what a scientific theory should be about.
 It is one of the merits of Bricmont's book that it includes a whole chapter on this issue, what it means to \emph{explain}
something in science, and why the lack of an objective explanation \emph{is} a serious scientific deficiency of QM.
 Thus, in chapter 3, ``Philosophical Intermezzo,'' Bricmont eloquently argues the case that ``the conceptual problems of
quantum mechanics are internal to the physical theory itself and do not have a `philosophical' solution'' (quoted from p.73).
 This chapter is simultaneously a call for rational understanding and
against positivistic, idealistic, or solipsistic attitudes, which were in vogue in scientific circles 
at the time QM was created, and which continue to influence the thinking of many physicists.

 While Bohr declared that such an objective explanation is impossible, 
 Einstein on the other hand was convinced that orthodox QM is an incomplete theory, and that a more complete theory 
which explains the quantum formalism in terms of objective physical processes is possible.
 Bricmont explains that Einstein favored the statistical ensemble interpretation of the quantum state, meaning the quantum 
state (the wave function $\Psi$) is not fundamental but rather representing (through Born's interpretation of $|\Psi|^2$)
something akin to a Gibbs ensemble of more fundamental state variables which had yet to be discovered.
 Instead of more neutrally calling them ``omitted,'' or ``missing variables,'' 
the name ``hidden variables'' was given to these; another unlucky choice of terminology which suggests that those
variables would be inaccessible --- metaphysics to those with a positivistic attitude that ``nothing is real unless
you measure it.''
 Whatever one calls these variables, 
 Albert Einstein, Boris Podolsky, and Nathan Rosen in their famous EPR paper \cite{EPR} concluded that the assumption of the correctness
of the statistical quantum mechanical predictions in concert with Einstein locality (i.e., ``no spooky action at a distance,''
as Einstein called it) implies that certain deeper ``elements of reality'' (read: hidden variables) 
must exist but are missing from QM. 
 As Bricmont explains in chapter 7, there is evidence that Einstein was trying to convey this to Bohr already in
1927, to no avail.

 By ``spooky action at a distance'' 
Einstein most definitely meant a violation of what his relativity theory postulated: an event $A$ can influence an
event $B$ if and only if $B$ lies in the forward lightcone of $A$. 
 In plain language, Einstein took it for granted that once the two particles of an originally close pair of particles
which ``acted in concert'' separated far enough spatially,
each particle will be ``on its own'' and not immediately disturbed (at least not in any essential way)
by acts of ``measurement'' of the other particle's state.
 Anything else is ``spooky.''

 But ``QM'' has earlier been said to stand for ``non-relativistic'' quantum mechanics, so 
how can one coherently combine ``non-relativistic'' QM predictions with ``relativistic notions of locality'' 
and expect to come to a compelling conclusion?
 Fair enough, and this brings us to chapter 4 and the second ``mystery.'' 

 In chapter 4.3 Bricmont explains what is at the bottom of Einstein's ``spooky action at a distance'' 
without running afoul of a naive mismatch of Lorentzian vs. Galilean relativistic structures. 
 Bricmont formulates four points, with a subdivision, which in the non-relativistic 
setting distinguishes ``spooky action at a distance'' from the usual ``Newtonian action at a distance'' 
(like Newtonian gravity, or Coulombian electricity).
 Basically, the distinction is that 
there is an instantaneous influence from one system on another which \emph{does not diminish with distance}, and 
furthermore it \emph{cannot be used to send signals} from one system to another.
 (There is also an ``individuation'' involved, stipulating that memory effects from earlier experiments could be
prevented from influencing the outcomes of present experiments.)

 In Newtonian physics, the change in position of a point particle with non-zero mass will instantaneously change the 
gravitational force it exerts on any other particle in the universe, but its strength diminishes with distance $r$ as 
$1/r^2$, so that very distant particles are ``essentially'' undisturbed by each other.
 Even then one could in principle use this tiny influence for instantaneous controlled signaling. (Bricmont 
explains that in a Newtonian world this would only be limited by technical abilities to detect tiny forces.)
 In this sense, what Einstein identified as ``spooky action at a distance'' is a distinctly different nonlocality
from Newton's ``fading-away but controllable action at a distance'' even in a ``non-relativistic'' theory. 
 (The issue of relativity is briefly commented upon in chapter 4 and taken up again in chapter~5.)

 After clarifying what Einstein meant by locality,
Bricmont explains in a crystal clear manner John Bell's proof that Einstein locality is violated by quantum physics.
 In a nutshell, the logic is this: 1. EPR show that postulating the correctness of certain QM predictions 
together with postulating Einstein locality implies that ``hidden variables'' exist that have certain particular
qualities; 2. Bell shows that postulating hidden variables with those particular qualities 
implies an inequality (the famous ``Bell's inequality'') 
which violates other predictions of QM. 
Conclusion: \emph{if QM makes the correct predictions, Einstein locality is false}!

  As Bricmont relates (in chapter 7), the simple logic of Bell's two-part argument is frequently
misunderstood by physicists.
 This was frustrating to Bell (quoted on p.265):\footnote{Essentially all quotations of
  John S. Bell in Bricmont's book are from the collection of Bell's works on the foundations of QM \cite{BellBOOKsec}.}
 ``It is remarkably difficult to get this point across, that determinism is not 
  a presupposition of the analysis.''\footnote{N.B.: By ``determinism'' Bell here refers to the hidden variables 
  whose qualities EPR concluded must (pre-)\emph{determine} the outcome of the experiments. 
  This becomes clear through the context.}
  The most common mistake is to ignore the EPR part of the argument and to insist that Bell did not prove that
QM is nonlocal but only that ``hidden variables do not exist,'' which therefore would demonstrate
``the non-existence
of microscopic reality.''\footnote{Bell's frustration is shared by others:
 ``I can't take it anymore. Eventually [somebody] will win the Nobel Prize in physics
  for having shown that reality does not exist.'' \cite{Detlef}\vspace{-15pt}}
 True, Bell proved that certain types of
 hidden variables cannot exist if QM makes empirically correct predictions.
 But it was the existence of precisely those types of hidden variables that EPR established by assuming certain QM
predictions \emph{and} Einstein locality.
  Bricmont is very explicit about it (p.124): ``[F]or Bell, his result, combined with the EPR argument, was not 
a `no hidden variables theorem,'  but a nonlocality theorem, the result about the impossibility of hidden variables 
being only one step in a two-step argument.''
 
 But there is another frequent objection to the claim that Bell proved that quantum physics is nonlocal,
typically coming from mathematical quantum field theorists who point out that ``quantum field theory is 
Lorentz-covariant and therefore local,'' so that ``Bell must be mistaken.''
 Bricmont, himself trained in constructive quantum field theory, and who states (p.172) that 
``The predictions of quantum field theory are extremely impressive,'' also has this to say (p.172): 
 ``... quantum field theory essentially allows us to compute {`scattering cross-sections'}.''
 [N.B.: This is done with the help of a so-called $S$-matrix which maps an asymptotically incoming state from 
(using relativists' terminology) ``past timelike (or null) infinity'' into an asymptotically outgoing state at 
``future timelike (or null) infinity'' (cf. \cite{GlimmJaffe}), but has nothing to say about what happens in between.] 
 ``The results are, it is true, invariant under Lorentz transformations,''
... ``but contrary to received opinion, it is not true that there exists a fully relativistic quantum theory. 
 After all, such a theory should be a generalization of non-relativistic quantum mechanics and should include 
(as one of its approximations) the results of EPR--Bell experiments (or any similar experiments involving a 
collapse of the quantum state when the latter is an entanglement of states whose parts are spatially separated). 
 But that would require a relativistic treatment of the collapse as a physical process, and such a treatment simply 
does not exist.
 Moreover, one would encounter enormous difficulties if one tried to formulate it,~...'' 
 And here is Bell himself (quoted on p.173): ``[The usual quantum] paradoxes are simply disposed of by the 
1952 theory of Bohm, leaving as the question, the question of Lorentz invariance. 
 So one of my missions in life is to get people to see that if they want to talk about the problems of quantum mechanics --- 
 the real problems of quantum mechanics --- they must be talking about Lorentz invariance.''
 (N.B.: By ``quantum mechanics'' Bell here really means ``quantum theory.'')

 The ``1952 theory of Bohm'' referred to above by Bell is explained in chapter~5.
 Bricmont calls it the de Broglie--Bohm theory, for 
it was Louis de Broglie who originally did suggest this theory at the 1927 Solvay conference \cite{deBroglieSOLVAY}.
 It seems that de Broglie had this theory in mind already when he wrote his doctoral dissertation, but at that
time he had no wave equation for his ``de Broglie waves'' --- Schr\"odinger's equation came two years later, inspired
by (but, curiously enough, also based on a misunderstanding of) de Broglie's proposal.
 Einstein, although initially encouraging de Broglie to continue along this line of research (see Bricmont's Einstein quote
at the bottom of p.237 / top of p.238 --- incidentally, de Broglie did not heed Einstein's advice for 25 ensuing years), 
after David Bohm rediscovered de Broglie's theory 25 years later \cite{Bohm52}, in a letter to Max Born (quoted on p. 270) wrote:
``[The de Broglie--Bohm] way seems too cheap to me.''\footnote{Einstein, to the best of the reviewers knowledge,
  didn't elaborate what was ``too cheap'' about the ``de Broglie--Bohm way'' to quantum mechanics.
  Einstein did, however, write down some concrete criticism of the de Broglie--Bohm theory in his contribution
 to the Festschrift on the occasion of Max Born's retirement from the Tait Chair of Natural Philosophy at the
 University of Edinburgh (Hafner Pub. Co. Inc., N.Y., 1953), Ref.[172] in Bricmont's book. 
 Remarkably enough, after Bohm refuted Einstein's criticism Einstein asked the editors that Bohm's refutation
be included also in the Festschrift --- and so it was.
 Basically, Bohm's argument was that Einstein had criticized the de Broglie--Bohm theory based on a particular
solution which is not physically observable, similarly to criticizing Newtonian mechanics by pointing out that
it allows (never observed) solutions showing a needle standing on its tip on a glass plate forever, without analyzing 
whether this is a stable (read: observable) situation.}
 Bricmont speculates that Einstein found the theory ``too cheap'' because it is manifestly nonlocal,  something he
could not accept (as Bricmont explained in chapter 4). 
 But, as Bricmont points out (p.271): ``Einstein was writing this 12 years before Bell's theorem.'' 

 The principles of the de Broglie--Bohm theory are very simple: In the non-relativistic version, there
is the same Galilean three-dimensional physical space 
and one-dimensional physical time 
as in Newton's point particle mechanical theory
of the physical world,\footnote{I ignore here that Newton wrote he thought that there \emph{is}  a preferred 
 ``absolute space,'' for his theory does not rely on this supposition.}
and in physical space there is matter which is made of a huge number $N$ of point particles which move through 
space as time goes on, in a deterministic manner (this idea is also shared with Newton's theory), 
\emph{but the law of motion is decidedly non-Newtonian}!
 The law of motion (in the spin-less version) says that the velocity of the $k$-th point particle is obtained by evaluating 
the $k$-th coordinate \emph{gradient of the phase $\Phi$ of the Schr\"odinger wave function} $\Psi = |\Psi|e^{i\Phi}$
at the \emph{actual} $N$ \emph{particle configuration}; the Schr\"odinger wave function in
turn is just the solution of QM's Schr\"odinger wave equation on $N$-particle configuration space.

 So here, in addition to the wave function $\Psi(t,\mathbf{x})$ with $\mathbf{x}\in\Rset^{3N}$ 
one does have an extra variable, the actual configuration $\mathbf{X}(t)\in\Rset^{3N}$, which is missing from the orthodox
theory. 
 The equation of (spin-less) motion for $\mathbf{X}(t)$ is the collection of ``guiding equations''\footnote{Incidentally, 
if in the guiding equation one sets $\hbar \Phi (t,\mathbf{x}) =: S(t,\mathbf{x})$,
it becomes identical to the one in the Hamilton--Jacobi formulation of Newtonian point mechanics, except that $S(t,\mathbf{x})$
here does \emph{not} satisfy the Hamilton--Jacobi PDE, but rather a PDE which differs from it by an additive term which
depends only on $|\Psi|$. 
 Since this $|\Psi|$-dependend term symbolically vanishes as $\hbar \to 0$, 
one sees that Newtonian point particle mechanics is the classical limit of the de Broglie--Bohm theory
whenever $S(t,\mathbf{x})$ converges to a regular function as $\hbar \to 0$.
 Hamilton--Jacobi theory is not addressed in Bricmont's book, who tries to keep the mathematics at a more elementary level.\vspace{-10pt}}
$\frac{d}{dt}\mathbf{X}_k(t) =\frac{\hbar}{m_k}\nabla_k \Phi(t,\mathbf{x})|^{}_{\mathbf{x}=\mathbf{X(t)}}$ for all $N$
three-dimensional components of $\mathbf{X}(t)$.
 Bricmont also explains how spin can easily be incorporated by replacing the $N$-body Schr\"odinger with the $N$-body
Pauli equation, and by rewriting the guiding equation as 
$\frac{d}{dt}\mathbf{X}(t) =\frac{\mathbf{j}(t,\mathbf{X}(t))}{\rho(t,\mathbf{X}(t))}$,
where ${\mathbf{j}(t,\mathbf{x})}$ and ${\rho(t,\mathbf{x})}$
are what in QM is usually called ``quantum probability current'' and ``quantum probability density,'' 
respectively, computed from the solution of either the Schr\"odinger equation (the spin-less case) or the
Pauli equation (when spin is included); note
that in the de Broglie--Bohm theory these quantities do not fundamentally have Born's statistical
(read: ensemble) meaning but a dynamical one.
 Note also that the de Broglie--Bohm theory does \emph{not} postulate special ``hidden spin variables'' to incorporate ``spin.''

 It is clear already what this theory says about the world: matter is made of point particles which move in a certain 
non-Newtonian way.
 There's nothing incomprehensible about that. 
 In particular, the theory does not suffer from a ``reality deficiency disorder.'' 
 Now, the theory's aim is to explain everything QM says about matter in terms of how these fundamental constituents of matter, 
the point particles, move, including in particular an explanation of what happens in a ``measurement.''
 The question whether it succeeds is answered by Bricmont with a resounding ``Yes!''.

 There are few expositions of the de Broglie--Bohm theory, for instance the book by Bohm and Hiley \cite{BoHi} where
it goes under the name Bohm gave it, ``Causal interpretation of QM,'' and the one by Holland \cite{Ho} on ``The quantum 
theory of motion,'' and the one by D\"urr and Teufel \cite{BMbook} on ``Bohmian mechanics,'' but there are also the 
collections of works by Bell \cite{BellBOOKsec} and by D\"urr, Goldstein, and Zangh\`{\i} \cite{DGZworks}. 
 Bricmont closely follows the exposition of the theory given in the works of Bell, and of D\"urr, Goldstein, Zangh\`{\i}, and
their collaborators.
 Of course, he doesn't merely repeat what others have written about the matter, but condenses its 
essence into the space of a single chapter. 

 After having explained what the problem is with orthodox QM, and how the de Broglie--Bohm theory overcomes this problem,
Bricmont in chapter 6 raises the question whether other, much more popular attempts to overcome the ``measurement problem of QM,''
offer an alternative.

\vspace{-5pt}
 Chapter 6.1 addresses some variations of the ``Many-Worlds Interpretation,'' conceived of by
Everett and advocated by (among others) Bryce DeWitt, David Deutsch, Lev Vaidman, and the philosopher David Wallace.
 (But, as Bricmont explains, also Erwin Schr\"odinger's early attempt, which he subsequently dropped, to
extract objective physics from his wave function is a version of a ``many-worlds'' theory.)
 These ``interpretations'' reject the ``Copenhagen postulate'' that the
unitary quantum evolution is ``\emph{fundamental only between measurements}'' (and to be replaced by a non-unitary 
``collapse of the wave function'' during ``measurement'') and try to account for ``the real physical world {`out there'~}''
in terms of a de facto ``Pure Wave Function Ontology,''\footnote{This means 
``the wave function \emph{is all there is}.''\vspace{-15pt}}
but (except for Schr\"odinger's) they are very \emph{vague} about what precisely is ``a world,'' of which they claim there 
is a proliferation.
 (Schr\"odinger didn't think of his as a ``many-worlds'' theory, for it
does have a matter ontology in our real physical space, but as Valia Allori and her collaborators put it
(p.208): ``the world, in that theory, is like a TV set that is not correctly tuned so that a mixture of
several channels is seen at any given same time'' [except that one doesn't simultaneously see many movies,
but many `evolving material realities.'])

 Chapter 6.2 addresses the ``Spontaneous Wave Function Collapse'' theories, 
in particular the Ghirardi--Rimini--Weber theory, which was followed up by Bell, made relativistic by Rodi Tumulka, 
and more recently contemplated also by Weinberg --- these are theories in which the unitary evolution of the quantum
state is replaced by a stochastic evolution which ``collapses'' the quantum state of a system every once in a while 
\emph{objectively} without invoking a ``measurement'' for that purpose; they replace the vague orthodox talk of
``collapse of the wave function'' by implenting this notion in a mathematically precise manner.
 In this connection Bricmont also mentions other approaches based on modifying the unitary Schr\"odinger evolution, 
not into a stochastic but into a nonlinear deterministic evolution, associated with the names of Roger Penrose and 
of Steven Weinberg, amongst others.
 The crucial distinction of these approaches is that their predictions differ from those of orthodox quantum theory. 
 This feature is welcomed by experimentalists, but so far no experimental discrepancy with the QM predictions has been 
discovered (within the error bounds). 

 Chapter 6.3 addresses the ``Decoherent Histories'' approach associated with 
Bob Griffiths, Roland Omn\`es, and Murray Gell-Mann -- Jim Hartle, amongst others.
 ``Decoherent Histories'' are also known as ``Consistent Histories,'' somewhat ironically for, as Bricmont reports,
it was demonstrated by Shelly Goldstein, and independently Bassi and Ghirardi, that the ``Consistent Histories''
interpretation yields logical inconsistencies (unless you're willing to dispute the meaning of the word ``and''). 

  Chapter 6.4 addresses ``QBism'' (``Quantum-Bayesianism''), which has its origin in the more recent 
\emph{application of QM} called ``quantum information theory'' (started by Richard Feynman \cite{FeyII}, exploiting 
``quantum entanglement,'' and 
capitalizing on the path-breaking papers by EPR \cite{EPR}, Schr\"odinger \cite{Erwin}, and Bell \cite{BellBOOKsec}), 
and which enjoys a growing following (associated with the names 
Charlie Bennett, Chris Fuchs, Asher Peres, David Mermin, ...).
 In brief, ``QBism'' seems to be based on the idea that ``fundamental insights into QM'' will be obtained by (paraphrasing Bricmont)
``turning the logic upside down,'' by postulating some information-theoretical axioms and hoping to obtain QM.
 Thus, in ``QBism'' QM is viewed as a subjective theory.
  Here is Chris Fuchs (quoted on p.225): ``~`Whose information?' `Mine!' Information about what? `The consequences (for me) of my 
actions upon the physical system!'  It's all `I-I-me-me mine,' as the Beatles sang.'' 
 Bricmont, himself an aficionado of the Bayesian (subjective) interpretation of ``probabilities,'' disputes that
QM (and by extension, physics) is \emph{fundamentally} about ``our subjective experiences'' of the world (p.226):
``QBists claim not to be solipsists (people who write articles and books to convince others of their views rarely 
 claim to be solipsists), but if physics is all about updating my subjective experiences, then in what sense does 
 it differ from solipsism? 
 Travis Norsen calls it solipsism FAPP\footnote{``For All Practical Purposes;'' the acronym is due to John Stewart Bell.}
 but it is rather solipsism in denial, since that approach does not deal with anything except one’s subjective experiences; 
 so the difference with solipsism is only that one denies being solipsist, without saying anything about what could exist 
 outside one’s conscious states of mind.''
 (Chapter 3 comes in handy here to appreciate what Bricmont is trying to convey!)
  
 Bricmont compares these approaches to the de Broglie--Bohm theory and, in chapter 6.5, summarizes his conclusions:
``[W]e claim to have shown here that [the dBB theory] is the only theory that is free from any of the following problems:
\begin{itemize}
\item Inconsistency, like the decoherent histories approach. \vspace{-10pt}
\item Making different predictions than quantum mechanics, hence also than the de
Broglie--Bohm theory, as the spontaneous collapse theories do, while adjusting its
parameters so that no empirical contradiction can be detected. \vspace{-10pt}
\item Putting the observer back at the center of our physical theories, as QBism does, or
as the defenders of the decoherent histories approach also do (sometimes). \vspace{-10pt}
\item Being ill-defined or hard to make sense of, like the many-worlds interpretation, or the `bare' GRW theory 
(the one without ... [an] ontology).\footnote{Read: ``without a clear statement of what \emph{exists}.''\vspace{-15pt}}''
\end{itemize}

 Bricmont's list of ``alternative'' proposals to overcome QM's ``measurement problem'' is certainly incomplete,
as the author himself points out, but it does compare the author's favorite (the dBB theory) with what appears
to be the currently most popular ``beyond-Copenhagen'' proposals, and he makes a compelling (if harsh) case for
the rational superiority of the de Broglie--Bohm theory.
 Some other well-known proposals, such as 
``~{`quantum logic,' `quantum probabilities,' or the `modal interpretation,'~}'' are left aside,
 ``partly because they are not that popular any more'' (p.199).
 Also J\"urg Fr\"ohlich's very recent 
``ETH\footnote{Standing, not for ``Eidgen\"ossische Technische 
Hochschule'' but, for ``Events, Trees, Histories.''}
 theory'' (see \cite{ETH}) is not addressed --- time will tell whether it will withstand the scrutiny of inquisitive peers.

 If the de Broglie--Bohm theory would itself have been an invention of the recent past, 
the book could have ended here, except for some technical appendices (I haven't yet mentioned any of these; chapters
 2--6 have appendices which supply some technical details to the explanations of the main sections).
 However, the theory has been around pretty much since the inception of QM. 
 And so Bricmont asks (p.228): ``If the de Broglie--Bohm theory has so many qualities, why is it so generally ignored?''
 and answers: ``To answer this, we must turn to the history of quantum mechanics and take into account some
 psychological and sociological factors.''
 Chapters 7 (``Revisiting the history of quantum mechanics'') and 8 (``Quantum mechanics and our {`culture'~}'') 
are devoted to this. 
 It is in these two chapters where Bricmont's long-time engagement in fighting unscientific irrationality, not only in the
post-modern but in particular also in the scientific communities, is once again on full display. 
 
 So, at the end of the day, does Bricmont succeed in `making sense of (non-relativistic) quantum mechanics'?
 Absolutely ``Yes!'', in the opinion of this reviewer.\footnote{
 Full disclosure: The reviewer is on record \cite{KieFoP} for expressing the opinion 
that ``the de Broglie--Bohm theory (a.k.a. Bohmian mechanics), which is presumably the simplest theory which 
explains the orthodox quantum mechanics formalism, has reached an exemplary state of conceptual clarity and 
mathematical integrity. No other theory of quantum mechanics comes even close.''}
 What's more, Bricmont inspires readers, in particular students and junior researchers,
to stop to think, and not to submit to the gospels of false prophets.
 In particular, if you think that ``quantum theory is incomplete because we haven't succeeded in
quantizing gravity yet,'' then think again!
 Near the end of his final chapter 8.4 (``Understanding quantum mechanics: an unfinished story''), Bricmont 
points out: ``A genuine merging of quantum mechanics and [not only general, but even already 
special] relativity is still an open problem, and might be called the great unrecognized frontier of physics.''

 I wish this book had been around already when I was a student. 
 It is a \emph{must read} for anyone who seriously cares about \emph{understanding} 
what our ``most fundamental scientific theories'' say about the physical world.

\footnotesize{

}
\end{document}